\documentstyle[twoside,fleqn,jhuwkshp,psfig]{article}
\newcommand{\etal}{{\it et al.} }
\newcommand{\asca}{{\it ASCA} }

\newcommand{\xmm}{{\it XMM-Newton} }
\newcommand{\chandra}{{\it Chandra} }
\newcommand{\hetg}{{\it HETG} }

\newcommand{\rxte}{{\it RXTE} }
\newcommand{\fekalfa}{${\rm Fe} \rm \ K\alpha$ }
\newcommand{\tartarus}{{\sc tartarus} } 
\newcommand{\mcg}{{MCG~$-$6-30-15} }
\newcommand{\fignfifty}{{Fig. 1} }
\newcommand{\fignlone}{{Fig. 2} }

\newcommand{\figmcg}{{Fig. 3} }
\newcommand{\figthreec}{{Fig. 4} }
\newcommand{\fignfourcont}{{Fig. 5} }

\begin{document}

\title{Subtleties in Measuring Iron K Lines in AGN}
 
\author{T. Yaqoob 
\address{Department of Physics and Astronomy,
Johns Hopkins University, 3400 N. Charles Street, Baltimore, MD 21218, USA.}
\address{NASA/GSFC, Laboratory for High Energy Astrophysics, Greenbelt, MD 20771, USA.} 
U. Padmanabhan $^{a}$, 
T. Dotani \address{Institute of Space and Astronautical Science, 3-1-1 Yoshinodai, Sagamihara, Kanagawa 229-8510,
Japan.}
I. M. George $^{b}$ \address{Joint Center for Astrophysics, University of Maryland, Baltimore County,1000 Hilltop Circle, Baltimore, MD21250.},
T. J. Turner $^{b}$  $^{d}$,
K. Weaver, $^{b}$,
K. Nandra $^{b}$ \address{Universities Space Research Association.}}

\begin{abstract}
\chandra and \xmm observations are showing that the \fekalfa emission
lines in type 1 AGN are composite, in general consisting of a narrow
and a broad component. We review the latest \chandra HETG results and
compare the line profiles with those measured by {\it ASCA}.
The narrow \fekalfa line components necessitate re-modeling of the line profiles
measured previously and revision of the parameters of the relativistic broad components,
as well re-interpretation of variability studies of the \fekalfa lines.
There has been concern in the literature that changes in the \asca 
calibration have made the \fekalfa lines narrower and weaker
than originally thought. 
We explicitly demonstrate the effect of changes in the \asca calibration
   since launch on the measurements of the \fekalfa lines in AGN. We find that 
   the
   differences in measured parameters (centroid energy, width and equivalent
   width) are insignificant. In particular both the intrinsic width and 
   equivalent width change by $\sim 8\%$ or less. 
   The reason for the recent claims that the changes
   in the \asca calibration have made the \fekalfa lines systematically
   narrower is due to the fact that in some cases the lines are complex
   and cannot be described by a single Gaussian. In such cases the current
   calibration models the narrow core only, when in fact the overall
   profile may be very broad. Thus the width of the line can be seriously
   underestimated if it is not modeled correctly. We point out that it is
   incorrect to compare different calibrations using models which do not
   describe the data. We present some new results for the highest S/N
   broad Fe K line profile in the entire \asca archive, from NGC 4151.
   We show that 
   the broadness of the iron line and its shape is not very sensitive to the
   precise form of the complex continuum, contrary to popular belief.

\end{abstract}

\maketitle

\section{The Complex Iron Lines in AGN}

\asca found that the  \fekalfa
fluorescent emission line in Seyfert~1 galaxies is
often very broad,
and this is generally interpreted as the result of an origin in
matter in an accretion disk rotating around a central black hole
(see \cite{Fabi2000} and references therein).
The line profile is thought to be sculpted by characteristic
gravitational and Doppler energy shifts.
Currently, study of the  \fekalfa emission line is
the only way to probe
matter within a few to tens of gravitational radii
in supermassive  black holes. Indeed, the broad \fekalfa lines provide
some of the strongest evidence to date for the existence of
black holes.
 
\begin{figure}[hbt] 
\centerline{\psfig{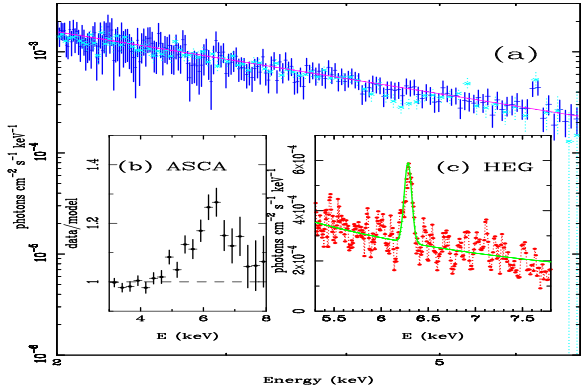}}
\vspace{-8mm}
\caption{\footnotesize (a) \chandra \hetg spectra for NGC 5548 from the HEG (solid error
bars) and
MEG (crosses with dotted error bars).
Solid line is the best-fitting power-law
continuum to HEG plus MEG data but with the best-fitting HEG
normalization. The
$\pm 1$ orders are summed.
The narrow  \fekalfa
emission-line component is clearly detected in both the HEG and MEG.
(b) \asca SIS0+SIS1 relativistic  \fekalfa line profile
for the July 1996 observation of
NGC 5548. (c)
Close-up of the HEG spectrum; the data
are smoothed with a boxcar  five bins wide, where the bin size is
$0.005 \AA$.
The solid line is the best-fitting model which
consists of a power-law continuum and a narrow Gaussian.
Energy scale in all three panels
is for the {\it observer's} frame.
}\label{fig:fignfifty}
\end{figure}

It has been known since the early days of \asca that
the \fekalfa line in AGN
is not always tremendously broad (e.g. \cite{Ptak1994}) and that the
profile comes in a variety of
shapes (e.g. \cite{Yaqo1996,Nand1997}).
It has also been known
that Seyfert type 1.9--2.0 galaxies have a predominantly narrow
\fekalfa line, with FWHM less than $\sim 10,000$ km/s, probably
originating in cold matter far from the black hole (e.g. NGC 2992;
\cite{Weav1996}). It was apparent that many Seyfert 1.5--1.9 galaxies
likely have composite narrow and broad \fekalfa line components
(e.g. \cite{Weav1993,Yaqo1995,Weav1997,Weav1998}).
\chandra and \xmm have now shown that composite narrow and
broad \fekalfa lines are common even in Seyfert 1 galaxies and
quasars \cite{Yaqo2001a,Kasp2001,Reev2000,Poun2001}.
\fignfifty (from \cite{Yaqo2001a})
shows the narrow line in the \chandra
HETG data for NGC~5548, compared to non-contemporaneous \asca data showing the
overall, broad \fekalfa line profile. \fignlone shows
a montage of more measurements of the \fekalfa narrow-line component
in type~1 AGN, again compared with non-contemporaneous \asca 
measurements. Note that \chandra doesn't always clearly detect the
broad component of the line. This is not surprising since the 
effective area of HETG in the Fe-K band is so small. {\it The HETG
data do not in any case reject the corresponding broad line measured
by \asca}. Also note that \chandra {\it does} detect a broad
feature in NGC~4051, F9, and NGC~4151 (for the latter see 
Weaver \& Yaqoob, these proceedings).
Of course we do need simultaneous measurements with
instruments that can measure both the narrow and broad components
and we have a program to do this with \chandra and \rxte (some with \xmm).
 
\begin{figure}[hbt]
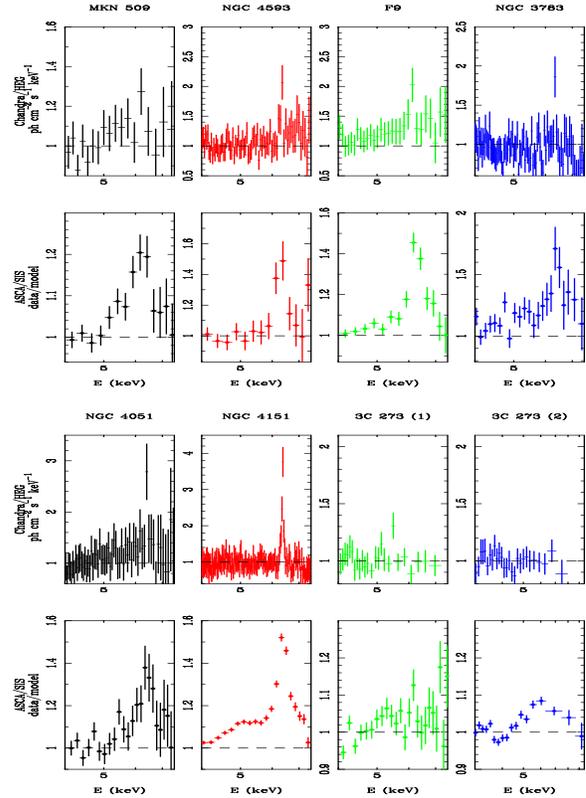
 
\centerline{\psfig{file=j_ty_fig2.ps,width=3.0in,height=2.0in}}
\vspace{3mm}
\centerline{\psfig{file=j_ty_fig3.ps,width=3.0in,height=2.0in}}
\vspace{-8mm} 
\caption{\footnotesize The Fe-K band as seen by the HEG for several
type~1 AGN, compared with the \fekalfa line profiles as seen
with non-contemporaneous \asca observations
(except for obs(1) of 3C~273 for which the \chandra and \asca data {\it are} simultaneous). 
A narrow core is common. In some cases
the HEG can actually detect the broad part of the \fekalfa line.
}\label{fig:fignlone}
\end{figure}                                                                             

The equivalent widths of the narrow lines (NL) range from tens of eV to
over a 100 eV in some cases. In all type~1 AGN studied by \hetg
so far (excluding NLS1), the narrow line appears to originate in
cold matter, with its center energy around 6.4 keV.
Only in one case (NGC~3783; \cite{Kasp2001}) is the narrow \fekalfa line resolved (in NGC~5548 it was marginally resolved \cite{Yaqo2001a}).
In that case its location is somewhere in between the BLR and NLR. In all the
other cases, upper limits on the FWHM are $<10,000$ km/s. 
Its location could
be any one or more of: BLR, NLR, obscuring torus, or even the outer regions of
the accretion disk. The latter is possible because {\it we do not actually know the
distribution of \fekalfa line emissivity over the disk}. The
distribution may be flatter than $r^{-2}$ and need not
even be a power-law. For example,
if the disk is illuminated by a distribution of magnetic
flares all over the surface then the line emissivity
could be very flat with radius.
It may then appear that there is a separate
NL, when in fact the disk is producing both the NL {\it and}
the broad line (BL).
If the spatial
distribution of flares varies with time, the NL will appear to vary.
Indeed, \fignlone shows that a narrow \fekalfa line is detected in one
observation of 3C~273 but not in another observation which has the same
continuum luminosity. This strongly argues for a disk origin of the narrow line
in 3C~273.
A program is under-way to retroactively re-interpret previous measurements
of the \fekalfa line profiles and variability for as many AGN as
possible, in the light of new \chandra and
\xmm results. 

\section{\fekalfa Lines and \asca Calibration}

Recently Lubi\'{n}ski and Zdziarski (\cite{Lubi2001} hereafter LZ01) have claimed that
the \fekalfa lines in AGN are {\it systematically} narrower and weaker
(smaller equivalent width, or EW) than had previously been thought.
They speculated that
this result can be attributed to changes in the calibration of the \asca
instruments. Specifically, LZ01 refer to the calibration used by
Tanaka \etal 1995 \cite{Tana1995} (hereafter T95) and Nandra \etal 1997 \cite{Nand1997} (hereafter N97) and  
that used in the \tartarus AGN database\footnote{http://tartarus.gsfc.nasa.gov/}.
We call the T95 and N97 calibration `OLD' and the \tartarus calibration `CURRENT'.
\figmcg compares the \fekalfa line profiles for MCG~$-$6$-$30$-$15 using
the actual data and calibration files used by T95 \cite{Tana1995} and {\sc tartarus}.
A similar comparison is shown
for 3C~120 in \figthreec but this time calibration effects
are isolated by using the same spectral files but constructing OLD and CURRENT
calibration for them.
Detailed spectral fit results and technical details 
on the different calibrations can be found in \cite{Yaqo2001b}.
Now these observations of MCG~$-$6$-$30$-$15 and 3C~120
have much higher signal-to-noise than typical $\sim 1$ day \asca observations
(which constitute the LZ01 sample), yet the calibration effects on the
\fekalfa line are small. In fact simulations show that the CURRENT
calibration reduces the 
EW and intrinsic width by only $\sim 8\%$.
Moreover, all of the observations considered by LZ01 were made
before radiation damage made a significant impact on the
CCDs.
Therefore the principal differences
between the OLD and CURRENT calibration for these early data
are, (1) improvements in the SIS response matrix, (2)
multiplicative, energy-dependent corrections for the effective area
(the so-called `arf filter'), and (3)
changes in the X-ray Telescope (XRT) ray-tracing (see \cite{Yaqo2001b} for details).
The latter is the {\it only} change that can make the
OLD and CURRENT calibration have different effects
on different observations since the first two are
independent of position on the detector and independent
of time of the observation
(as long as that time is early enough in the mission).
Even the change in XRT response is a
smooth function of source off-axis angle, and its variation
is small given the limited range of off-axis angles
in observations. {\it Changes in calibration cannot therefore
produce wild differences in the \fekalfa lines in some sources
and small differences in others as LZ01 claim.}
If one examines the \fekalfa line profiles
for the type~1 AGN in LZ01 which do give wildly different line widths
compared to N97, one finds that the lines claimed to be narrowest
by LZ01 are in fact some of the broadest. However, the lines are
can be complex and cannot always be modeled by single Gaussians,
so the LZ01 spectral fits could model the narrow 
line core only, grossly underestimating the width and
EW. 

\begin{figure}[thb] 
\centerline{\psfig{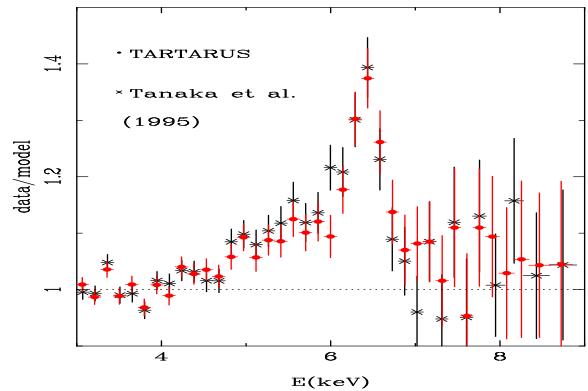}}
\vspace{-8mm}
\caption{\footnotesize Effect of changes in the \asca calibration on
the \fekalfa profile for \mcg (\asca AO2),
shown for co-added SIS0 and SIS1 data (see text).
Crosses correspond to the Tanaka \etal data \cite{Tana1995} (OLD calibration) and
filled circles correspond to \tartarus data (CURRENT calibration).
}\label{fig:figmcg}
\end{figure}

\begin{figure}[bht] 
\centerline{\psfig{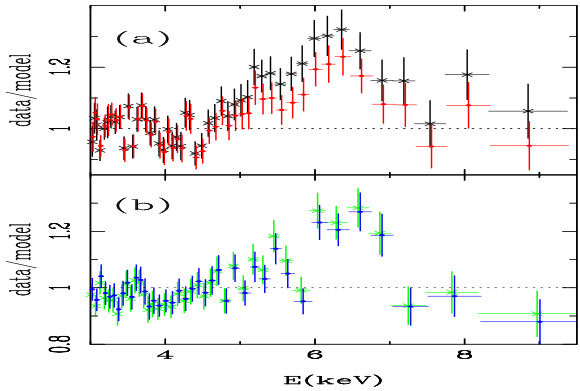}}
\vspace{-8mm}
\caption{\footnotesize Effect of changes in the \asca calibration on
the \fekalfa profiles for 3C~120 (\asca AO6),
shown separately for (a) SIS0 and (b) SIS1. Here calibration effects are
isolated by using the same spectra.
Crosses correspond to the OLD calibration and
filled circles correspond to the CURRENT calibration.
}\label{fig:\figthreec}
\end{figure}

\vspace{-1cm}
\section{The Broad \fekalfa Line in NGC 4151}

NGC~4151 has been observed by \asca several times. 
An unprecedented AO8 $\sim 13$ day observation in May 2000 
(exposure $\sim 330$~ks) yielded
an \fekalfa 
line profile with {\it the highest signal-to-noise 
} for any AGN in the entire \asca archive.
The 5--7 keV band in the \asca SIS0 AO8 NGC~4151 data
contains $\sim 6.5$ times more photons
(in total) than MCG~$-$6$-$30$-$15.
 
In \fignfourcont 
we show the result of fitting the 0.6--10 keV \asca AO8 (SIS0) data for
NGC~4151 with two different complex continuum models (power law plus
a dual cold
absorber and a warm absorber, as well as a thermal emission component)
and double Gaussians for the \fekalfa line. It can be seen that the resulting
\fekalfa line profile is hardly sensitive to the precise form of the
continuum. In fact even if we fit the 3--10 keV band only with a simple
single absorber (plus power-law), the \fekalfa line profile is not
substantially affected. 
It is also important to note that {\it the relative strength of
the red wing is variable} in general: this has been known
for MCG~~$-$6$-$30$-$15 for some time. The AO8 data clearly
show this for NGC~4151 but that work will be presented elsewhere.
In any case, we should not be surprised, if during a particular
observation of NGC~4151 the red wing appears to be weak.

\begin{figure}[htb]
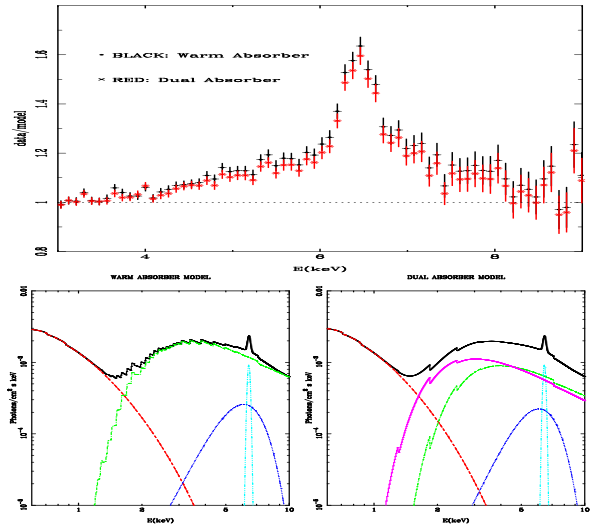
 
\centerline{\psfig{file=j_ty_fig6.ps,width=3.0in,height=1.4in}}
\centerline{\psfig{file=j_ty_fig7.ps,width=1.5in,height=1.3in} \psfig{file=j_ty_fig8.ps,width=1.5in,height=1.3in}}
\vspace{-8mm}
\caption{\footnotesize Insensitivity of NGC~4151 \fekalfa line profile to detailed
continuum form (see text). 
}\label{fig:fignfourcont}
\end{figure}

The statistical quality of the \asca AO8 data is good enough to
reject claims that the broad \fekalfa line in NGC~4151 is really
the result of a thick absorber mimicking the broad line. Considering
a  spectral fit
to the SIS0 data in the 0.6--10 keV
range with a power law plus warm absorber, thin thermal continuum,
narrow \fekalfa line, plus a very thick, partially covering,
absorber,
to model the
apparent broad red wing of the \fekalfa  line, we obtain a poor fit.  
Compared to a similar
model with the thick absorber replaced by a broad Gaussian
the difference in $\chi^{2}$ for the thick absorber model compared
to the broad line model is 43, rejecting the former
model
at an extremely high level of confidence. The d.o.f.
for the two models are equal because the thick absorber model
has free covering factor and photon index.


\vspace{-10mm}

\normalsize

\section*{ACKNOWLEDGMENTS}
We acknowledge support from the following grants: NCC5-447 (TY, UP),
NAG5-10769, NAS8-39073 (TY),
NAG5-7067, NAG5-7385 (KN,TJT). 

\small
\begin{thebibliography}{19}

\bibitem{Fabi2000} Fabian, A. C., \etal 2000, PASP, 112, 1145

\bibitem{Ptak1994} Ptak, A., Yaqoob, T., Serlemitsos, P. J., \etal 1994, ApJ, 436, L31

\bibitem{Yaqo1996} Yaqoob, T., Serlemitsos, P. J., Turner, T. J., \etal 1996, ApJ, 470, L27

\bibitem{Nand1997} Nandra, K., George, I. M., Mushotzky, R. F., \etal 1997, ApJ, 476, 70 (N97)

\bibitem{Weav1996} Weaver, K. A., Nousek., J., Yaqoob, T., \etal 1996, ApJ, 458, 160

\bibitem{Weav1993} Weaver, K. A., \etal 1993, ApJ, 423, 621 

\bibitem{Yaqo1995} Yaqoob, T., Edelson, R.,  Weaver, K. A., \etal 1995, ApJ, 453, L81

\bibitem{Weav1997} Weaver, K. A., Yaqoob, T., Mushotzky, R. F., \etal 1997, ApJ, 474, 675. 

\bibitem{Weav1998} Weaver, K. A., \& Reynolds, C. S. 1998, ApJ, 503, L39   

\bibitem{Yaqo2001a} Yaqoob, T. \etal 2001, ApJ, 546, 759

\bibitem{Kasp2001} Kaspi, S. 2001, these proceedings 

\bibitem{Reev2000} Reeves, J., \etal 2000, A\&A, 365, L134 

\bibitem{Poun2001} Pounds, K. A., \etal 2001, ApJ, 559, 181 


\bibitem{Lubi2001} Lubi\'{n}ski, P., \& Zdziarski, A. A. 2001, 323, L37  

\bibitem{Tana1995} Tanaka, Y., \etal 1995, Nature, 375, 659 

\bibitem{Yaqo2001b} Yaqoob, T. \etal 2001, \\  
{\tt http://heasarc.gsfc.nasa.gov/docs/ \\
asca/calibration/oldvscurrent.html}



\end{thebibliography}
\end{document}